\documentclass[prl,twocolumn,showpacs, amsmath,amssymb,aps]{revtex4-1}
\usepackage{graphicx}
\usepackage{bm}

\begin{document}
\newcommand{\balpha}{\mbox{\boldmath$\alpha$}}

\title{Multipolar interference for non-reciprocal nonlinear generation}

\author{Ekaterina Poutrina$^{1,2,^*}$ and Augustine Urbas$^{1}$}

\affiliation{$^{1}$Materials and Manufacturing Directorate, Air Force Research Laboratory, Wright Patterson Air Force Base, Ohio 45433, United States}
\affiliation{$^{2}$ UES, Inc., 4401 Dayto-Xenia Road, Dayton, OH 45432}
\email{ekaterina.poutrina.ctr.ru@us.af.mil}

\date{\today}

\begin{abstract}
We show that nonlinear multipolar interference allows achieving not only unidirectional, but also
a non-reciprocal nonlinear generation from a nanoelement, with the direction of the nonlinearly
produced light decoupled from that of at least one or several of the excitation beams. Alternatively, it may
allow inhibiting the specified nonlinear response in a nanoelement or in its periodic arrangement
by reversing the direction of one of the pumps. The described phenomena exploit the fact that,
contrary to the linear response case, nonlinear magneto-electric interference stems from a combination of additive and multiplicative processes and includes an interference between various terms within the electric and magnetic partial waves themselves. We demonstrate the introduced concept numerically using an example of a plasmonic dimer geometry with realistic material parameters.
\end{abstract}


\maketitle
Nonlinear optical antennas have been actively discussed as a tool for enhancing the efficiency of nonlinear response, while allowing for the polarization and directionality control of nonlinear generation on the nanoscale~\cite{Nov2003,Kaur_PRL,Nature_2008_HG_gener,PRL_2009_single_gold,NL_core_shell, Nov_enhancing, Maier_NanoLett, Maier_ACSnano, Poutrina_FWM, Nov_Cohcontrol, Nov_OA, NL_antennas_review}. Such efficient interchange between enhanced, sub-diffraction localized near field and directional, nonlinearly generated far field can serve as a keystone in applications in optical sensing, integrated optic circuits, or quantum information processing. The nanoscale nature of the interaction, however, prevents regular mechanisms such as phase matching from having an impact on the directionality of the response. As a result, the interference between the multipolar modes produced on scattering by the nanoelement~\cite{Kerker, Engheta_opt_theorem, OL_opt_theorem}--the process inherent to nanoscale interactions-- is at the heart of the directionality control. Indeed, similar to the linear response case~\cite{VH_unidir_science, Kuznestov,Novotny_linear, Kivshar_ACS_nano,VH_SRR}, the interference between the nonlinearly produced electric and magnetic dipolar modes has been discussed theoretically~\cite{Alec_theory} and demonstrated experimentally~\cite{Alec_exp} to produce a unidirectional nonlinear generation in the microwave regime from deeply sub-wavelength metamaterial layers. Nonlinear multipolar response has been also actively discussed in relation to optical nanoantennas~\cite{Kaur_multipole_2008, PRL_octupole, Kaur_OE_dip_lim, Kaur_NJP_multipole, Kivshar_Scherb_THG, Kivshar_Scherb_NLinterferenceTHG, Kivshar_NL_sandwitch}. Here, we show that nonlinear multipolar interference can not only, similar to the linear regime, ensure a unidirectional nonlinear generation from nanoelements, but also offers capabilities fundamentally different from those provided by its linear counterpart. For example, a single nanoelement, as well as a metasurface formed by a planar arrangement of such nonlinear optical antennas, can exhibit a non-reciprocal nonlinear generation; or ensure a directionally selective inhibition of the nonlinear response for certain respective directions of the fundamental beams.

Consider a nanoelement in which a set of electric and magnetic dipolar modes is induced along the $x$ and $y$ Cartesian axes, respectively, via some optical response to radiation incident along $z$ direction, and assume no other multipoles acquire a significant strength within the same spectral range. Radiation from such orthogonal, spatially-superimposed electric and magnetic dipoles results in the following angular distribution of the $E_\theta$ far field component of the electric field scattered in the $xz$ ($\varphi = 0)$ plane~\cite{Poutrina_FB, SM1}, with angles $\theta$ and $\varphi$ in a standard spherical coordinate:
\begin{equation}
\left. E_{\theta } \right|_{\varphi =0} =\frac{k^2}{4\pi }\frac{e^{ikr} }{r} \left[\frac{1}{\varepsilon _{0} } p_{x0} \cos \theta +\eta m_{y0} \right].
\label{eqn1}
\end{equation}
Here, $p_{x0}$ ($m_{y0}$) denotes the amplitude of the $x$ ($y$) component of the induced electric (magnetic) dipole, $\eta\equiv\sqrt{\mu_0/\varepsilon_0}$ is the vacuum impedance, with $\varepsilon_0$ and $\mu_0$ being the permittivity and the permeability of vacuum, $r$ is the length of the radius-vector directed from the scatterer (located at the origin) to the observation point in the far field, and $k$ is the wavenumber. One can show $E_{\varphi}=0 $ in the $\varphi = 0$ scattering plane, and the radial component $E_r$ falls off in the far field.

As follows from Eq.~1, the radiation from the above set of dipoles can be interferometrically suppressed in either backward ($\theta = 180^o$) or forward ($\theta = 0^o$)  direction (subject to the optical theorem limitations~\cite{Engheta_opt_theorem, OL_opt_theorem, Poutrina_FB}) when matching the strengths of the electric and magnetic dipolar modes either with the same or with the opposite signs, respectively $\left(1/\varepsilon _{0}p_{x} = \pm\eta m_{y}\right)$. Such suppression can clearly take place independently of whether the dipolar modes are induced through a linear or through a nonlinear interaction. In the case of a nonlinear response, however, each of the dipolar moments can result from a sum of several terms~\cite{Alec_theory, Fiebeg},  the number of terms being, in general, dependent on the order of the nonlinear process. For the second order response at frequency $\omega_3 =\omega_1+\omega_2$, assuming a monochromatic excitation, the relations can be written as follows:
\begin{subequations}
\begin{flalign}
\bf{p}_0\left( {{\omega _3}} \right)& \scalebox{1}{$= \frac{\epsilon_0}{\sqrt{I}}\left[\eta^2\alpha _{emm}^{\left( 2 \right)}\left( {{\omega _3};{\omega _1},{\omega _2}} \right):{\bf{H}_0}\left( {{\omega _1}} \right){\bf{H}_0}\left( {{\omega _2}} \right)\right.$} \nonumber\\
&\scalebox{1}{$+\left. \alpha _{eee}^{\left( 2 \right)}\left( {{\omega _3};{\omega _1},{\omega _2}} \right):{\bf{E}_0}\left( {{\omega _1}} \right){\bf{E}_0}\left( {{\omega _2}} \right)\right]$}\nonumber\\
&\scalebox{1}{$+ \frac{\epsilon_0}{\sqrt{I}}\left[\eta\alpha _{eem}^{\left( 2 \right)}\left( {{\omega _3};{\omega _1},{\omega _2}} \right):{\bf{E}_0}\left( {{\omega _1}} \right){\bf{H}_0}\left( {{\omega _2}} \right)\right. $}\nonumber\\
&\scalebox{1}{$+\left. \eta\alpha _{eme}^{\left( 2 \right)}\left( {{\omega _3};{\omega _1},{\omega _2}} \right):{\bf{H}_0}\left( {{\omega _1}} \right){\bf{E}_0}\left( {{\omega _2}} \right)\right]$},\\
{\bf{m}_0}\left( {{\omega _3}} \right)& \scalebox{1}{$= \frac{1}{\sqrt{I}}\left[\alpha _{mem}^{\left( 2 \right)}\left( {{\omega _3};{\omega _1},{\omega _2}} \right):{\bf{E}_0}\left( {{\omega _1}} \right){\bf{H}_0}\left( {{\omega _2}} \right)\nonumber\right. $}\\
&\scalebox{1}{$\left. + \alpha _{mme}^{\left( 2 \right)}\left( {{\omega _3};{\omega _1},{\omega _2}} \right):{\bf{H}_0}\left( {{\omega _1}} \right){\bf{E}_0}\left( {{\omega _2}} \right)\right]$}\nonumber\\
&\scalebox{1}{$+ \frac{1}{\sqrt{I}}\left[ \frac{1}{\eta}\alpha _{mee}^{\left( 2 \right)}\left( {{\omega _3};{\omega _1},{\omega _2}} \right):{\bf{E}_0}\left( {{\omega _1}} \right){\bf{E}_0}\left( {{\omega _2}} \right)\right. $}\nonumber\\
&\scalebox{1}{$\left. +\eta\alpha _{mmm}^{\left( 2 \right)}\left( {{\omega _3};{\omega _1},{\omega _2}} \right):{\bf{H}_0}\left( {{\omega _1}} \right){\bf{H}_0}\left( {{\omega _2}} \right)\right]$};
\end{flalign}
\end{subequations}
where  $\textbf{E}_0$ and $\textbf{H}_0$ are the vector amplitudes of the incident field, the time-harmonic factor been omitted, $I\equiv \left|\bf{E}_{0} \left(\omega_1\right)\right| \left|\bf{E}_{0}\left(\omega_2\right)\right|$,  $\textbf{p}_0$ and $\textbf{m}_0$ are vectors of the nonlinearly induced electric and magnetic dipole moments, and the double-dot product denotes the summation over all Cartesian components. The subscripts $``e"$ and $``m"$ indicate the type (electric or magnetic) of the dipolar modes participating in the interaction, with the second and the third subscripts denoting the types of the contributing fundamental modes at $\omega_1$ and $\omega_2$, respectively, and the first one the type of the produced dipolar mode at $\omega_3$. We emphasize that, noting a pure electric intrinsic response in the case of non-magnetic materials assumed here, the coefficients on the right hand side of Eq.~2 represent the effective, rather than the intrinsic hyperpolarizabilities, with the origin of the nonlinear response being different for plasmonic or other nanoelements made of centro-symmetric materials~\cite{Dadap_NL_origin_1, Dadap_NL_origin_2, Sipe_hydrodin, Shen_NL_origin, Scalora_hydrodin, Ciraci_hydrodin_1, Ciraci_hydrodin_2} versus those possessing an intrinsic second-order nonlinear response. The normalization is chosen to keep the single units of $[$m$^3]$ for all hyperpolarizabilities types, and frequency dependence follows conventional notation~\cite{Boyd}.

One can think of the various terms entering the right hand sides of Eqs. 2 and 3 as of different pathways leading to the certain type (electric or magnetic) of the nonlinear response. Note that the terms in the first and the second brackets in each of the Eqs.~2(a) and 2(b) comprise the sets of the effective hyperpolarizabilities of two different types, polar and  axial~\cite{Birss, Symmetry_and_magnetism, Alec_theory}, respectively.

Consider now a situation where only the first term in each of the Eqs.~2(a) and 2(b) is non-zero and assume $\alpha^{(2)}_{emm, xyy} = \pm\alpha^{(2)}_{mem, yxy}$, with the rest of the Cartesian components negligible. According to Eq.~2, these terms share the same pathway through the magnetic field of the fundamental beam at $\omega_2$. As a result, switching the phase of just the magnetic vector of this fundamental wave (an event occurring on reversing the propagation direction of that beam) will produce a simultaneous phase change in \textit{both} (nonlinearly generated) electric and magnetic dipolar modes. This results in a \textit{simultaneous} sign change before both terms in Eq.~1. According to that equation, the direction of the interferometric suppression and doubling of the nonlinearly generated field will be then preserved with respect to a fixed laboratory coordinate system, making the generation "non-reciprocal" with respect to the direction of that fundamental wave. This scenario is qualitatively different from linear scattering, where the pathways are always different for electric and magnetic responses, hence reversing the direction of incidence necessarily switches the directions of interferometric suppression and doubling of the scattered field~\cite{Poutrina_FB, Poutrina_multipolar}.

A similar non-reciprocity in the nonlinear response will occur for any other combination of the effective nonlinear susceptibilities of a similar strength that share the same, either electric or magnetic, pathway (e.g., the condition \scalebox{0.92}{$\alpha^{(2)}_{eem,xxy} = \alpha^{(2)}_{mee,yxx}$}, with the rest of hyperpolarizabilities negligible, ensures preservation of generation direction when reversing the fundamental wave at $\omega_1$). The non-reciprocity with respect to \textit{all} fundamental waves at once, however, requires the equality of hyperpolarizabilities belonging to different symmetry groups (e.g., \scalebox{0.92}{$\alpha^{(2)}_{emm} = \alpha^{(2)}_{mmm}$} ), and thus would be forbidden in elements of a single group symmetry. The principle can be extended to higher-order nonlinear processes, as well as to higher order multipolar modes.
\begin{figure}[tbp]
\centering\includegraphics[width=\columnwidth]{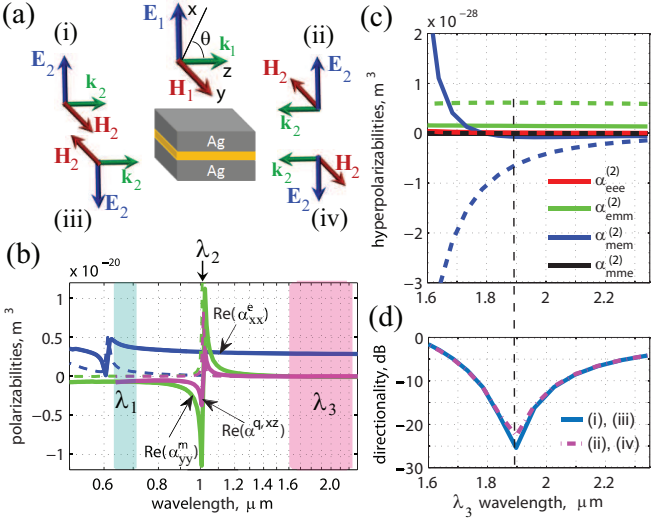} \caption{(a) The four pump arrangements used in Fig.~2. Dimer geometry: 120x120x35 nm silver~\cite{JC} strips with the 11 nm thick BBO~\cite{BBO} spacer; $\chi^{(2)}_{xxx} = 2.3$~pm/V~\cite{BBO_NL}. (b) Real (\textit{solid}) and imaginary (\textit{dashed}) parts of the retrieved linear electric ($\alpha^e_{xx}$) and magnetic ($\alpha^m_{yy}$) dipolar and the electric quadrupolar ($\alpha^{q,xz}$) polarizabilities; $\lambda_1$ and $\lambda_2$ indicate the wavelengths used to produce the DFG response at $\lambda_3$  in (c) and (d). (c) Retrieved real (\textit{solid}) and imaginary (\textit{dashed}) parts of the effective hyperpolarizabilities. (d) Directionality \textit{D} attains peak in the negative $z$ direction at 1.9~$\mu$m, where $\alpha^{(2)}_{emm} = -\alpha^{(2)}_{mem}$ with the rest of the hyperpolarizabilities being negligible, demonstrating a strongly directional and non-reciprocal, with respect to the pump at $\omega_2$, nonlinear generation. A small difference between the forward ((i),(iii)) and the reverse ((ii), (iv)) excitations is due to the nonzero $\alpha_{eee}$ term which does not share the same ``pathway" through the magnetic field of the wave at $\omega_2$ with the $\alpha^{(2)}_{emm}$ and $\alpha^{(2)}_{mem}$ terms; hence, it changes sign differently when reversing that pump direction, modifying the resulting electric and magnetic dipolar modes (Eq.~2). Finite elements method incorporated in COMSOL Multiphysics software package (www.comsol.com) is used for the numerical analysis.}
\end{figure}

While an option of such a non-reciprocity in the direction of nonlinear generation is unique by itself and can be employed in a variety of applications, manifestation of effective electromagnetic hyperpolarizabilities and nonlinear magneto-electric interference can result in other phenomena unavailable with natural nonlinear media. Each (electric or magnetic) mode by itself can result from the interference effects between the various terms entering the right hand sides of Eqs.~2(a) and 2(b). Note that the terms within a single symmetry group in either electric or magnetic mode (terms in each square brackets) necessarily depend on the different types of the fundamental field vector of a given frequency.  Consequently, reversing the direction of one of the fundamental waves will produce a sign change before \textit{only one} of the two interfering terms in each bracket and the nonlinear dipolar responses will be different for co- and counter-propagating pumps. Further, if, for example, a geometry/spectral position can be found ensuring the equality of \textit{all} the terms within one of the symmetry groups (e.g., \scalebox{0.92}{$\alpha^{(2)}_{eee} = \alpha^{(2)}_{emm} = \alpha^{(2)}_{mem} = \alpha^{(2)}_{mme}$}, with the rest terms zero), the nonlinear generation will be, according to Eq.~1, unidirectional for co-propagating pumps, but inhibited for counter-propagating pumps.
\begin{figure}[tbp]
\centering\includegraphics[width=\columnwidth]{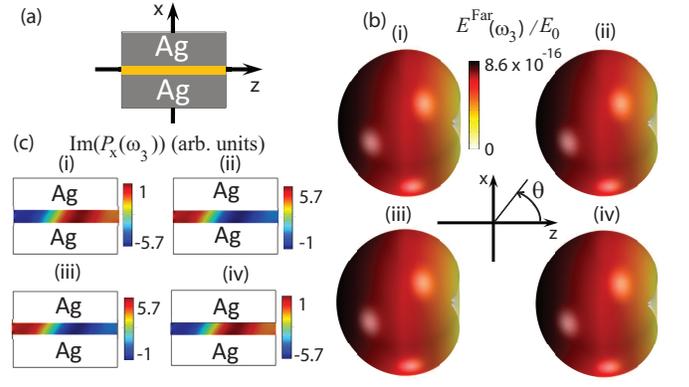} \caption{(a) Geometry cross section shown in (c). (b) Nonlinearly generated far field  for the four pump arrangements from Fig.~1(a) at $\lambda_3 = 1.9$~$\mu$m where the only non-negligible hyperpolarizabilities are $\alpha^{(2)}_{emm} = -\alpha^{(2)}_{mem}$. Due to sharing the same ``pathway" through the magnetic field of the pump at $\omega_2$ (Eq.~2), both (nonlinearly induced) electric and magnetic dipolar modes change sign  \textit{simultaneously} on reversing the direction of the magnetic field vector of the fundamental wave at $\omega_2$ (sets (ii) and (iii)); while no change, compared to (i), occurs in (iv). According to Eq.~1, the direction of the nonlinear generation is thus preserved in all four arrangements. Some difference between the forward and the reverse excitations is due to the small nonzero $\alpha^{(2)}_{eee}$ term. (c) Dominant component of the nonlinear polarization within the spacer layer at 1.9~$\mu$m, for the same four pump arrangements. Although the direction of nonlinear generation is preserved with all four arrangements, the induced nonlinear polarization is of the opposite sign in both the (ii) and the (iii) sets, with respect to that in (i) and (iv), following the phase of the magnetic vector of the fundamental wave at $\omega_2$, even though purely electric intrinsic second-order susceptibility is assumed. Location of Ag stripes in shown schematically for clarity. }
\end{figure}

To verify the feasibility of observing the above phenomena with physical nanoelements, we consider the process of difference frequency generation (DFG) in plasmonic dimers formed by two silver strips separated by a thin layer of dielectric (Fig.~1(a)). The choice of the nonlinear process is aimed at ensuring that the nanoelement is small compared to the nonlinearly generated wavelength, to minimize the manifestation of higher order multipoles in the nonlinearly generated field. Our preliminary hydrodynamic model~\cite{Sipe_hydrodin, Scalora_hydrodin, Ciraci_hydrodin_1, Ciraci_hydrodin_2} analysis has shown that the main contribution to the second-order nonlinear response in such a geometry results from the nonlinearity of the spacer material, in line with the recent results on the third order response in similar structures~\cite{Ciraci_THG}. The nonlinear response due to the metal has therefore been neglected in further analysis. Furthermore, the simulations have shown that the major contribution to the nonlinear response comes from the $\chi^{(2)}_{xxx}$ component of the intrinsic second-order nonlinear susceptibility tensor, with other tensor components producing a negligible impact even though accounted for. This is expected due to the dominance (up to two orders of magnitude) of the local fundamental electric field component orthogonal to the spacer layer within the spacer volume~\cite{Poutrina_FB, Poutrina_multipolar}. We thus choose the $\chi^{(2)}_{iii}$ tensor component of the BBO spacer to be along the $x$ axis.

Linear polarizabilities for the dimer geometry of Fig.~1(a), retrieved using our previously developed procedure~\cite{Poutrina_multipolar}, are shown in Fig.~1(b).  The same procedure can be used to identify the dipolar and quadrupolar partial waves in the nonlinearly generated field. One can repeat the procedure for four combinations of fundamental beams directions~\cite{SM}, arriving at the set of expressions for the effective nonlinear polarizabilities of the following form:
\begin{subequations}
\begin{flalign}
&\scalebox{1}{$\alpha _{emm,xyy}^{\left( 2 \right)} = \frac{0.25}{\epsilon_0\sqrt{\textrm{I}}}\left(p_{3x}^{+,+}-p_{3x}^{+,-}-p_{3x}^{-,+}+p_{3x}^{-,-}\right)$},\\
&\scalebox{0.98}{$\alpha _{mem,yxy}^{\left( 2 \right)} = \frac{\eta}{4\sqrt{\textrm{I}}}\left(m_{3y}^{+,+}-m_{3y}^{+,-}+m_{3y}^{-,+}-m_{3y}^{-,-}\right)$},
\end{flalign}
\end{subequations}
with similar expressions holding for the rest of the eight hyperpolarizabilities entering Eq.~(2)~\cite{SM}. In Eq.~3, the $p_{3x}^{+,-}$ ($m_{3y}^{+,-}$) and similar terms indicate the amplitude of the $x$ ($y$) component of the electric (magnetic) dipolar mode generated at frequency $\omega_3$ retrieved for a certain combination of forward ($``+"$) and reverse ($``-"$) directions of the fundamental beams. The first (second) superscript refers  to the direction of the wave at $\omega_1$ ($\omega_2$). The resulting hyperpolarizabilities obtained when placing the frequency $\omega_2$ at the "magnetic" resonance peak, while sweeping $\omega_1$ (Fig.~1(b)), are presented in Fig.~1(c). We verified the accuracy of both linear and nonlinear retrievals by ensuring a very good agreement between the numerical and the analytical far field angular distributions (the latter obtained using the retrieved dipolar and quadrupolar amplitudes)~\cite{SM}. Note that, although allowed in the retrieval, all axial hyperpolarizabilities vanish nearly identically, indicating the predominant polar nature of the dimer geometry.

As seen in Fig.1(c), \scalebox{0.9}{$\alpha^{(2)}_{emm}=-\alpha^{(2)}_{mem}$} and the rest of the hyperpolarizabilities are either identically zero or negligible near 1.9~$\mu$m, reproducing precisely the situation discussed earlier. Assuming the beam at $\omega_1$ propagating in the positive $z$ direction (Fig.~1(a)), one can expect the nonlinear generation near 1.9~$\mu$m to be interferometrically suppressed in a hemisphere encompassing the negative $z$ semi-axis, independently of the (forward or reverse) direction of the beam at $\omega_2$.

The analysis of the nonlinearly generated far field confirms the above prediction (Figs.~1(d) and 2(b)), with the directionality ratio $D\equiv 20\log_{10}\left(\left|E_3^{\textmd{Far}}\right|\mid_{\theta = 0^o}/\left|E_3^{\textmd{Far}}\right|\mid_{\theta = 180^o}\right)$ attaining the (negative) 26~dB peak at 1.9~$\mu$m for all four pump arrangements shown Fig.~1(a). Note that due to the placement of one of the pumps nearly precisely at the ``magnetic" resonance peak where the real part of the linear magnetic dipolar (as well as of the electric quadrupolar) polarizability almost vanishes, the real parts of the hyperpolarizabilities are negligible near 1.9~$\mu$m, and it is matching their imaginary parts that ensures the unidirectionality of the nonlinear generation. While matching the imaginary parts with the opposite signs is not surprising noting the presence of (nonlinear) gain, it is worth noting that the limitation on backscattering imposed by the optical theorem can be thus lifted in the process of nonlinear generation, similar to the linear response case in the presence of gain~\cite{footnote, OL_opt_theorem}. Also, we find that for the dimer geometry the generation direction is independent overall from the direction of the fundamental beam at $\omega_2$ within the $xz$ plane, due to the dominance of the local fundamental electric field component normal to the spacer layer for all dimer orientations within the $xz$ plane~\cite{Poutrina_FB}, except for some variation in the nonlinear efficiency for excitation at an angle with respect to the \textit{z} axis. The latter assumes a similar non-reciprocity when using not only plane wave, but also line current sources.

Shown in Fig.~2(c) is the $x$ component of the nonlinear polarization $P_x(\omega_3)\equiv const \cdot \chi^{(2)}_{xij}:E_i(\omega_1)E_j(\omega_2)$ ($i, j$ standing for $x, y, z$), induced at 1.9~$\mu$m. Due to the placement of $\omega_2$ at the ``magnetic" resonance where the electric dipolar mode is weak, the phase of the majority of the \textit{local} fundamental electric field $E_i(\omega_2)$, and hence the sign of the nonlinear polarization, follow that of the \textit{magnetic} vector of the incident field of the pump wave at $\omega_2$.  As a result, the induced intrinsic nonlinear polarization is of the opposite phase in either of the sets (ii) and (iii), with respect to that in (i) and (iv). This, in turn, transfers to the \textit{simultaneously} negative phases of the nonlinearly induced electric and magnetic dipolar modes (both related to the volume integral of the intrinsic polarization~\cite{Jackson, Kaur_gaus}) in either (ii) or (iii) cases, and thus, as discussed earlier, to the preserved direction of energy generation in all (i)-(iv) sets. The situation is similar to the ``negative index" media where the direction of energy propagation is preserved when reversing simultaneously the phase velocities of both the electric and the magnetic partial waves~\cite{Veselago, Smith_PRL_neg_index_theory, Smith_Science_neg_index_exp}, only here the simultaneous phase reversal is achieved through a nonlinear interaction.

In conclusion, we have shown that the involvement of multiplicative processes in nonlinear multipole interference allows achieving not only directional, but also non-reciprocal (with respect to the direction of one or several pump beams) nonlinear generation from nanoscale objects. We developed the procedure for the retrieval of nonlinear magneto-electric polarizabilities and demonstrated numerically that the described phenomena can be achieved in the optical frequency range using a dimer geometry with realistic materials parameters. The geometry is not unique and can be optimized further for fabrication purposes. We predicted that interference between various hyperpolarisability terms within nonlinearly generated multipolar modes themselves can allow inhibiting a certain nonlinear process in a subwavelength unit designed to have the same strength of all symmetry-allowed types of its hyperpolarizabilities, while reversing the direction of one of the pumps would switch it back on. The latter phenomena might be especially useful in the radio frequency range where nonlinear response can be strong but is often undesirable in applications. In addition to the manifestation of multipolar responses of opposite parities but of comparable strengths, the described phenomena require a certain relation between various hyperpolarizability terms within each (nonlinearly generated) multipolar mode, and thus are not expected to be available in natural nonlinear media, but can be achieved via the effective nonlinear multipolar response of nanostructures. Combined with strong enhancements of the nonlinear response offered by plasmonic metasurfaces, the described phenomena can serve as a keystone in a variety of applications in integrated optics and photonics.
\acknowledgements
The Authors would like to thank Sergey Tretyakov for the detailed discussion of bi-anisotropy in the linear response case.

\vspace{-4mm}
{\small


\begin{thebibliography}{}
\bibitem{Nov2003} A. Bouhelier, M. Beversluis, A. Hartschuh, and L. Novotny, Phys. Rev. Lett. \textbf{90}, 013903 (2003).
\bibitem{Kaur_PRL} S.~Kujala, B.~ K.~Canfield, and M.~Kauranen, Y. Svirko, and J. Turunen, Phys. Rev. Lett. \textbf{98}, 167403 (2007).
\bibitem{Nature_2008_HG_gener} S.~Kim, J.~Jin, Y.-J.~Kim, I.-Y.~Park, Y.~Kim, and S.-W. Kim, Nature \textbf{453} 757 (2008).
\bibitem{PRL_2009_single_gold} T.~Hanke, G.~Krauss, D.~Tr\"{a}utlein, B.~Wild, R.~Bratschitsch, and A.~Leitenstorfer, Phys. Rev. Lett. \textbf{103}, 257404 (2009).
\bibitem{NL_core_shell} Y.~Pu, R.~ Grange, C.-L.~Hsieh, and D.~Psaltis, Phys. Rev. Lett. \textbf{104}, 207402 (2010).
\bibitem{Nov_enhancing} H.~Harutyunyan, G.~Volpe, R.~Quidant, and L.~Novotny, Phys. Rev. Lett. \textbf{108}, 217403 (2012).
\bibitem{Maier_NanoLett} H.~Aouani, M.~Navarro-Cia, M.~Rahmani, T.~P.~H.~Sidiropoulos, M.~Hong, R.~F.~Oulton, and S.~ A.~Maier, Nano Lett. \textbf{12}, 4997 (2012).
\bibitem{Maier_ACSnano} M.~Navarro-Cia and S.~A.~Maier, ACS Nano \textbf{6}, 3537 (2012).
\bibitem{Nov_Cohcontrol} S. G. Rodrigo, H. Harutyunyan, and L. Novotny, Phys. Rev. Lett. \textbf{110}, 177405 (2013).
\bibitem{Poutrina_FWM} E.~Poutrina , C.~Cirac\`{\i}, D.~J.~Gauthier, and D.~R.~Smith, Opt. Express \textbf{20}, 11005 (2012).
\bibitem{Nov_OA} H. Harutyunyan1, G. Volpe and L. Novonty, \textit{Optical antennas} (Cambridge University Press, 2013), chap.8.
\bibitem{NL_antennas_review} S.~B.~Hasan, F.~Lederer, and C.~Rockstuh, Mater. Today \textbf{17}, 478 (2014), and references therein.

\bibitem{Kerker} M. Kerker, D.-S. Wang, and C. L. Giles, J. Opt. Soc. Am. \textbf{73}, 765 (1983).
\bibitem{Engheta_opt_theorem} A. Al\`{u} and N. Engheta, J. Nanophotonics \textbf{4}, 041590 (2010).
\bibitem{OL_opt_theorem} B. Garc\'{\i}\'{y}a-C\'{a}mara, R. A. de la Osa, J. M. Saiz, F. Gonz\'{a}lez, and F. Moreno, Opt. Lett. \textbf{36}, 728 (2011).
\bibitem{VH_unidir_science} A.~G.~Curto,  G.~Volpe, T.~H.~Taminiau, M.~P.~Kreuzer, R.~Quidant, and N.~F. van Hulst, Science \textbf{329}, 930 (2010).
\bibitem{Kuznestov} Y. H. Fu, A. I. Kuznetsov, A. E. Miroshnichenko, Y. F. Yu, and B. Lukyanchuk, Nat. Commun. 4, 1527 (2013).
\bibitem{Novotny_linear} S.~Person, M.~Jain, Z.~Lapin, J.~Jose Saìenz, G.~Wicks, and L.~Novotny, Nano Lett. \textbf{13}, 1806 (2013).
\bibitem{Kivshar_ACS_nano} W. Liu, A. E. Miroshnichenko, D. N. Neshev, and Y. S. Kivshar, ACS Nano 6, 5489–5497 (2012).
\bibitem{VH_SRR}    I.~M.~Hancu, A.~G.~Curto, M.~Castro-L\'{o´}pez, M.~Kuttge, and N.~F.~van~Hulst, Nano Lett. \textbf{14}, 166 (2014).
\bibitem{Alec_theory} A.~Rose, S.~Larouche, E.~Poutrina, and D.~R.~Smith, Phys. Rev. A \textbf{86}, 033816 (2012).
\bibitem{Alec_exp} A.~Rose, D.~Huang, and D.~R.~Smith, Phys. Rev. Lett. \textbf{110}, 063901 (2013).
\bibitem{Kaur_multipole_2008} S.~Kujala, B.~K.~Canfield, M.~Kauranen, Y.~Svirko, and J.~Turunen, Opt. Express \textbf{16}, 17196 (2008).
\bibitem{PRL_octupole} J. Butet, G. Bachelier, I. Russier-Antoine, C. Jonin, E. Benichou, and P.-F. Brevet, Phys. Rev. Lett. \textbf{105}, 077401 (2010).
\bibitem{Kaur_OE_dip_lim} R.~Czaplicki, M.~Zdanowicz, K.~Koskinen, J.~Laukkanen, M.~Kuittinen, and M.~Kauranen, Opt. Express \textbf{19}, 26866 (2011).
\bibitem{Kaur_NJP_multipole} M. Zdanowicz, S. Kujala, H. Husu, and M. Kauranen, New J. Phys. \textbf{13}, 023025 (2011).
\bibitem{Kivshar_Scherb_THG} M.~R.~Shcherbakov, D.~N.~Neshev, B.~Hopkins, A.~S.~Shorokhov, I.~Staude, E.~V.~Melik-Gaykazyan, M.~Decker, A.~A.~Ezhov, A.~E.~Miroshnichenko, I.~Brener, A.~A.~Fedyanin, and Y.~S.~Kivshar, Nano Lett. \textbf{14}, 6488 (2014).
\bibitem{Kivshar_Scherb_NLinterferenceTHG} M.~R.~Shcherbakov, A.~S.~Shorokhov, D.~N.~Neshev, B.~Hopkins, I.~Staude, E.~V.~Melik-Gaykazyan, A.~A.~Ezhov, A.~E.~Miroshnichenko, I.~Brener, A.~A.~Fedyanin, and Y.~S.~Kivshar, ACS Photonics \textbf{2}, 578 (2015).
\bibitem{Kivshar_NL_sandwitch} S. Kruk , M. Weismann, A. Yu. Bykov , E. A. Mamonov, I. A. Kolmychek, T. Murzina, N. C. Panoiu, D. N. Neshev, and Y. S. Kivshar, ACS Photonics \textbf{2}, 1007 (2015).
\bibitem{Poutrina_FB} E.~Poutrina, A.~Rose, D.~Brown, A.~Urbas, and D.~R.~Smith, Opt. Express \textbf{21}, 31138  (2013).
\bibitem{SM1} See Supplemental Material at [URL will be inserted by publisher] for the additional visualisation of coordinates orientation.
\bibitem{Fiebeg} M. Fiebig, D. Fro¨hlich, B. B. Krichevtsov, and R.V. Pisarev, Phys. Rev. Lett. \textbf{73}, 2127 (1994).
\bibitem{Sipe_hydrodin} J. E. Sipe, V. C. Y. So, M. Fukui, and G. I. Stegeman, Phys. Rev. B \textbf{21}, 4389 (1980).
\bibitem{Dadap_NL_origin_1} J.~I. Dadap,  J. Shan, K.~B. Eisenthal, and T.~F. Heinz, Phys. Rev. Lett. 83, 4045 (1999).
\bibitem{Dadap_NL_origin_2} J.~I. Dadap, J.~Shan, and T.~F.~Heinz, J. Opt. Soc. America Bv \textbf{21}, 1328 (2004).
\bibitem{Shen_NL_origin} P. Guyot-Sionnest, W. Chen, and Y. R. Shen, Phys. Rev. B 33, 8254 (1986).
\bibitem{Scalora_hydrodin} M. Scalora, M. A. Vincenti, D. de Ceglia, V. Roppo, M. Centini, N. Akozbek, and M. J. Bloemer, Phys. Rev. A \textbf{82}, 043828 (2010).
\bibitem{Ciraci_hydrodin_1} C. Cirac\`{\i}, E.~Poutrina, M.~Scalora, and D.~R.~Smith, Phys. Rev. B 85, 201403 (2012).
\bibitem{Ciraci_hydrodin_2} C. Cirac\`{\i}, E.~Poutrina, M.~Scalora, and D.~R.~Smith, Phys. Rev. B 86, 115451 (2012).
\bibitem{Boyd} R. W. Boyd, \textit{Nonlinear Optics} (Academic, 2008).
\bibitem{Birss} R. Birss, Rep. Prog. Phys. \textbf{26}, 307 (1963).
\bibitem{Symmetry_and_magnetism} R.~R.~Birss, \textit{Symmetry and magnetism} (North-Holland Pub. Co., 1964).
\bibitem{Poutrina_multipolar} E.~Poutrina and A.~Urbas, J. Opt. \textbf{16}, 114005 (2014).
\bibitem{JC} P.~B.~Johnson and R.~W.~Christy, Phys. Rev. B \textbf{6}, 4370 (1972).
\bibitem{BBO} D. Eimerl, L. Davis, S. Welsko, E.K. Graham, and A. Zalkin,  J. Appl. Phys. \textbf{62}, 1968 (1987)
\bibitem{BBO_NL} V.~G.~Dmitriev, G.~G.~Gurzadyan, amd D.~N.~Nikogorasyan \textit{Handbook of Nonlinear Optical Crystals} (Springer, 1997).
\bibitem{Ciraci_THG} C.~Ciraci, M.~Scalora, and D.~R.~Smith, Phys. Rev. B \textbf{91}, 205403 (2015).
\bibitem{SM} See Supplemental Material at [URL will be inserted by publisher] for the derivation and the accuracy verification.
\bibitem{footnote} Indeed, as seen from Eq.~1, a perfect backscattering requires matching electric and magnetic dipolar moments with the opposite signs, making such situation impossible for passive elements for which the imaginary parts of multipolar contributions always stay positive. The presence of gain lifts this limitation.
\bibitem{Jackson} J. D. Jackson, \textit{Classical Electrodynamics} , Ch. 9, J. Wiley and Sons, Inc., (1998).
\bibitem{Kaur_gaus} M. J. Huttunen, J. M. Akitalo, G. Bautista, and M. Kauranen, New J. Phys. \textbf{14}, 113005 (2012).

\bibitem{Veselago} Veselago, V. G, Soviet Physics Uspekhi \textbf{10}, 509 (1968).
\bibitem{Smith_PRL_neg_index_theory} D.~R.~Smith, W.~J.~Padilla, D.~C.~Vier, S.~C.~Nemat-Nasser, and S.~Schultz, Phys. Rev. Lett. \textbf{84}, 4184 (2000).
\bibitem{Smith_Science_neg_index_exp} R.~A.~Shelby, D.~R.~Smith, and S.~Shultz, Science \textbf{292}, 77 (2001).


\end{thebibliography}
\end{document}